\documentclass[superscriptaddress,prb,twocolumn]{revtex4}%
\usepackage{graphicx}
\usepackage{amsmath}
\usepackage{color}

\begin{document}
\title{Surface-induced magnetism in C-doped SnO$_{2}$}

\begin{center}
\textcolor{blue}{APPLIED PHYSICS LETTERS \textbf{96}, 052508 (2010)} 
\end{center}

\author{Gul Rahman}
\email[]{grnphysics@yahoo.com}
\affiliation{Graduate Institute of Ferrous Technology, Pohang University of Science and Technology, Pohang 790-834, Republic of Korea}

\author{V\'{\i}ctor M. Garc\'{\i}a-Su\'arez}
\email[]{v.garcia-suarez@lancaster.ac.uk}
\affiliation{Departamento de F\'{\i}sica, Universidad de Oviedo \&
CINN, 33007 Oviedo, Spain}
\affiliation{Department of Physics, Lancaster University,
Lancaster, LA1 4YB, United Kingdom}

\begin{abstract}
The magnetism of C-doped SnO$_{2}$ (001) surfaces is studied using first-principles calculations. It is found that carbon does not induce magnetism in bulk SnO$_{2}$ when located at the oxygen site, but shows a large magnetic moment at the SnO$_{2}$ (001) surface. The magnetic moment is mainly contributed by the carbon atoms due to empty minority spins of $p$ orbitals and is localized at the surface and subsurface atoms. No magnetism is observed when the carbon atom is located at the subsurface oxygen sites. The origin of magnetism is discussed in the context of surface bonding.
\end{abstract}


\maketitle
In search of spintronic materials, many  materials, which are non-magnetic but can show magnetism with transition-metal (TM) doping\cite{gul1,gul2}, have been discovered. The best example is GaAs, which is a semiconductor, but exibits  interesting magnetic properties when doped with Mn.\cite{ref-Mn} The search for spintronic materials has also produced  magnetic compounds with transition temperatures ($T_\mathrm{C}$) higher than room temperature (RT). Among these high $T_\mathrm{C}$ compounds, a promising but less studied material is TM doped-SnO$_{2}$, which evoked a lot of attention when S. B. Ogale \textit{et.al}\cite{Ogale} found a giant magnetic moment (GMM) in Co-doped SnO$_{2}$. Following this discovery, TM doped-SnO$_{2}$ has been extensively studied experimentally~\cite{exp1,exp2,exp3} and theoretically.~\cite{theory1, theory2,theory3,theory4}

Very recently,\cite{Rahman} we investigated the origin of GMM in doped SnO$_{2}$. Our calculations showed that Sn vacancies are responsible for magnetism in this material. This allowed us to classify SnO$_{2}$ into a  class of magnetic materials where magnetism is made possible without magnetic impurities. Therefore, intrinsic defects in SnO$_{2}$ can play an important role to explain magnetism or even optical conductivity.\cite{alex} Recent theoretical calculations further showed that magnetism can be induced with nonmagnetic impurities.\cite{N-doped,K-doped} A good example of non-magnetic impurity is carbon, which has been recently predicted to induce magnetism in bulk materials.\cite{TiO2,CdS,ZnS}

All previous investigations focused mainly on bulk magnetism. However, there are no reports on the surface magnetism of SnO$_{2}$. The questions that will be addressed here are: what is the role played by the surface and is it possible to induce magnetism at the surface without TM impurities? We, therefore, go beyond bulk magnetism and show that a surface can also induce magnetism in SnO$_{2}$ without TM impurities, which proves that impurities that are non-magnetic in bulk can become magnetic on the surface. This is not a general rule because the electronic and magnetic properties depend strongly on the combination of host and impurity\cite{NZnO,ZnS,NSnO2} (e.g. other sp elements are magnetic in bulk SnO$_2$ \cite{N-doped,K-doped,NSnO2}) but demonstrates the feasibility of such effect.

\begin{figure}
\begin{center}
\includegraphics[width=0.35\columnwidth]{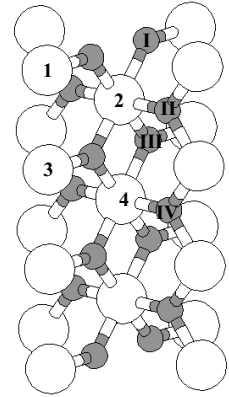}
\caption{Stoichiometric supercell of SnO$_{2}$ used in the calculations for the (001) surface. Big and small balls represent tin and oxygen atoms, respectively. Surface oxygen and tin atoms are represented by O(I) and Sn(1), respectively. The immediate subsurface atoms are represented by O(II) and Sn(2). Note that O(II) is surrounded by three Sn atoms, Sn(2), Sn(3), and Sn(4). The surface O(I) is surrounded only by two Sn atoms, Sn(1) and Sn(2). }
\label{crystal}
\end{center}
\end{figure}

Bulk SnO$_{2}$ has a tetragonal unit cell with two SnO$_{2}$ units (layers) per cell. Each Sn atom is coordinated to six O atoms situated at the vortices of a distorted octahedron. Each O atom is coordinated to three Sn atoms (labeled Sn(2), Sn(3), and Sn(4) in Fig.~\ref{crystal}) and all three O-Sn bonds lie on the same plane.

To study the bulk properties of C-doped SnO$_{2}$, a 2${\times2\times2}$ supercell was considered and a C atom was located at the O site. For surface magnetism, we considered a (001) surface of SnO$_{2}$ and used different numbers of layers. We constructed free stoichiometric slabs with total compositions of Sn$_{6}$O$_{12}$(six layers), Sn$_{7}$O$_{14}$ (seven layers, see Fig.~\ref{crystal}), and Sn$_{10}$O$_{20}$ (ten layers) separated by a vacuum region of 10\,\AA. Note that six layers are already enough to converge the SnO$_{2}$ surface.\cite{yuhua,Oviedo} To include the C atom different O sites, i.e., O(I), O(II), O(III), and O(IV), were considered for all layers. We will only focus on two cases, O(I) and O(II), because O(III) and O(IV) are similar to O(II).

First-principles calculations based on density functional theory within the local spin density approximation \cite{lsda} were performed using the
{SIESTA} code \cite{siesta} with a double-zeta polarized basis set. The Hamiltonian matrix elements were calculated on a real space grid defined with a plane-wave energy cutoff of $400$ Ry. A ${5\times 5\times 1}$ Monkhorst-Pack mesh was used for \textbf{k}-point sampling. All atomic positions were fully relaxed until all atomic forces were smaller than 0.05 eV/\AA.  For all calculations, the optimized lattice parameters \cite{Rahman} were used.

\begin{figure}
\begin{center}
\includegraphics[width=0.8\columnwidth]{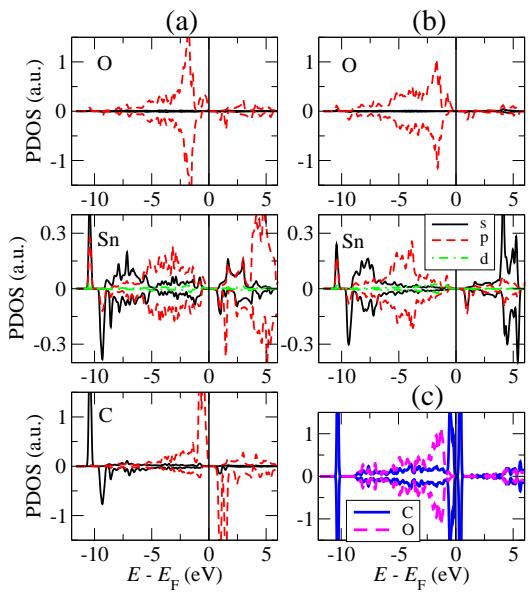}
\caption{PDOS on the surface (a) and subsurface (b) atoms when the C atom is located on the surface. Solid, dashed and dashed-dotted lines represent $s$,  $p$  and $d$ orbitals, respectively. The positive (negative) PDOS shows majority (minority) spin states. Panel (c) shows the PDOS of C and the nearest O in bulk C-doped SnO$_{2}$.}
\label{CaseoneDOS}
\end{center}
\end{figure}

Non-magnetic (NM) and spin-polarized (SP) calculations were performed for bulk C-doped SnO$_{2}$. The calculated atom projected density of states (PDOS) of C and O are shown in panel (c) of Fig.~\ref{CaseoneDOS}. The bulk PDOS shows no exchange splitting, which signals that C does not induce magnetism in bulk SnO$_{2}$. We found however that N can induce magnetism in bulk SnO$_{2}$, in agreement with previous studies.\cite{NSnO2} It was also confirmed that the O vacancy at the surface\cite{Trani} and in bulk\cite{Rahman} does not show any magnetism.

Figure~\ref{CaseoneDOS}(a) shows the PDOS on the orbitals of the surface atoms when the C atom is located at the surface O position, i.e., at O (I). Clearly, the C atom induces magnetism at the (001) surface of SnO$_{2}$. The low lying $s$ orbitals of C are spin-polarized and strongly hybridized with the surface $sp$ orbitals of Sn. The Fermi energy ($E_\mathrm{F}$) is mainly dominated by the $p$ orbitals of C, which indicates that magnetism is mainly induced by the $p$ orbitals and localized at the C atom. Majority $p$ spin states of the C atom are completely occupied and minority spin states are nearly empty, leading to a significant spin splitting. The impurity bands introduced by the C atom have then a large weight and are rather localized below and above the valence band maximum so that $E_\mathrm{F}$ is just between the majority and minority spin states(Fig~\ref{CaseoneDOS}(a)). Similar behavior was also observed in other C-doped bulk systems.\cite{TiO2,ZnS,ZnO} Surface states near $E_\mathrm{F}$ at the Sn and O atoms which have strong bonding with the C surface atom can also be seen.

The calculated magnetic moment is $2.00$ $\mu_\mathrm{B}$/cell. The C atom contributes about $1.75$ $\mu_\mathrm{B}$, according to the Mulliken populations given by SIESTA. The induced magnetic moments at the surface Sn(1) and O(I) atoms are $0.002$ and $0.077$ $\mu_\mathrm{B}$, respectively. The magnetic moments of the atoms in the slab have the same direction, indicating that there is a FM coupling between the C atom and the neighboring surface atoms. The PDOS on the subsurface atoms is shown in Fig.~\ref{CaseoneDOS}(b). The subsurface atoms (O(II), Sn(2)) are also polarized and contribute partially to magnetism. The induced magnetic moments at the subsurface Sn(2) and both the O(II) atoms are $0.002$$\mu_\mathrm{B}$ and $0.068$$\mu_\mathrm{B}$, respectively. In both cases (surface and subsurface), the O atoms have larger magnetic moments than the Sn atoms because their $p$ orbitals have larger weight near $E_\mathrm{F}$, where the C spin-split $p$ orbitals are localized.
\begin{figure}
\begin{center}
\includegraphics[width=0.8\columnwidth]{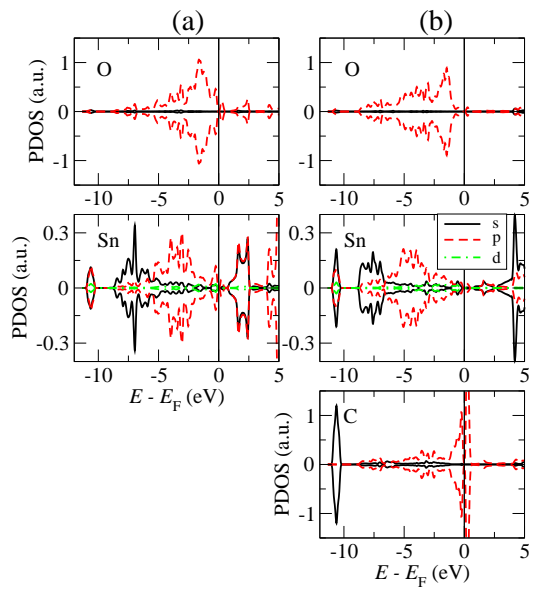}
\caption{PDOS on the surface (a) and subsurface (b) atoms when the C atom is located below the surface. Solid, dashed and dashed-dotted dotted lines represent $s$ and $p$ and $d$ orbitals, respectively. The positive (negative) PDOS shows majority (minority) spin states.}
\label{CasetwoDOS}
\end{center}
\end{figure}

We discuss now the case where the C atom is located at the subsurface O site, i.e., at O(II); this configuration resembles the case where the C atom is covered by layers of SnO$_{2}$, which is NM. Now the surface layer SnO$_{2}$ screens the C atom and therefore the system is expected to behave like a bulk C-doped SnO$_{2}$. This case does not produce magnetism (Fig.~\ref{CasetwoDOS}) since the exchange splitting in the $s$ as well as in the $p$ orbitals of the C atom is absent. The PDOS shows that the (001) surface has small surface states just below $E_\mathrm{F}$, which is in agreement with previous theoretical calculations.\cite{yuhua} A small peak, which is induced by the subsurface C $p$ orbitals just above $E_\mathrm{F}$, can also be seen. The PDOS also shows hybridization between the $p$ orbitals of C, Sn, and O atoms, particularly near $E_\mathrm{F}$. We can then conclude that C does not induce magnetism when located at the subsurface O(II) positions due to the lack of empty minority $p$ spin states, as opposed to the surface O(I) positions.

Figure~\ref{crystal} illustrates that each O atom is surrounded by three Sn atoms and each atomic layer consists of O-Sn-O units which have a total of 16 valence electrons. A simple electron counting suggests that when an O atom is replaced by a C atom, the total number of valence electron changes to 14 and hence induces holes in the system. At the (001) surface, the O atoms are not surrounded by three Sn atoms but by two Sn atoms. Hence, magnetism at SnO$_{2}$ (001) in this case is driven simultaneously by a surface effect (low coordination) and unpaired electrons. Calculations with two C located at the O sites show that the magnetic moment increases, suggesting that hole doping enhances magnetism. In bulk C-doped SnO$_{2}$ the C atom bonds directly to the three Sn atoms, which allow the 6 valence electrons to delocalize on the states associated to the bonds and reduce their Coulomb repulsion, leaving empty the localized state of the C atom, as can be seen in Fig.~\ref{CaseoneDOS} (c) and Fig.~\ref{CasetwoDOS} (b). In the surface, however, the number of bonds is reduced to two, which leaves two localized states, as can be seen in Fig.~\ref{CaseoneDOS} (a). The two electrons which are not associated to the bonds follow then Hund's rules to reduce again their Coulomb repulsion and occupy each state with parallel spins, which gives a total magnetic moment of $2.00$ $\mu_\mathrm{B}$/C.

Finally, similar calculations were also performed for different layers of SnO$_{2}$ and we always found that the C atom can only induce magnetism at the surface layers and the magnetic moment is only localized at the surface and subsurface atoms.

In summary, first-principles calculations showed that C induces a large magnetic momment at the SnO$_{2}$ (001) surface. The magnetic moment, which is localized at the surface and subsurface atoms, was mainly contributed by the C atom and partially by the O atoms. When the C atom was located at the subsurface O sites it did not show magnetism. The possible origin of magnetism was discussed in terms of surface bonding. Further work would be necessary to investigate similar phenomena at other semiconductor surfaces. We hope our results stimulate further experimental verification.

VMGS thanks the Spanish Ministerio de Ciencia e Innovaci\'on for a Juan de la Cierva fellowship and the Marie Curie European ITNs FUNMOLS and NANOCTM for funding.

\end{document}